\documentclass{INTERSPEECH2023}
\interspeechcameraready 
\usepackage[english]{babel}

\usepackage{soul}
\usepackage{amsmath}
\usepackage{graphicx}
\newcommand{\eg}{e.\,g.,\,}
\newcommand{\ie}{i.\,e.,\,}

\newcommand{\et}{{et al.\,}}

\title{U-DiT TTS: U-Diffusion Vision Transformer for Text-to-Speech}
\name{Xin Jing$^1$, Yi Chang$^3$, Zijiang Yang$^1$, Jiangjian Xie$^2$,\\Andreas Triantafyllopoulos$^1$, Bj\"orn W.\ Schuller$^{1,3}$}
\address{
  $^1$Chair of Embedded Intelligence for Health Care \& Wellbeing, University of Augsburg, Germany\\
  $^2$ School of Technology, Beijing Forestry University\\
  $^3$GLAM -- Group on Language, Audio, \& Music, Imperial College London, UK}
\email{xin.jing@informatik.uni-augsburg.de}

\begin{document}
\maketitle

\begin{abstract}

Deep learning has led to considerable advances in text-to-speech synthesis.
Most recently, the adoption of Score-based Generative Models~(SGMs), also known as Diffusion Probabilistic Models~(DPMs), has gained traction due to their ability to produce high-quality synthesized neural speech in neural speech synthesis systems. 
In SGMs, the U-Net architecture and its variants have long dominated as the backbone since its first successful adoption. 
In this research, we mainly focus on the neural network in diffusion-model-based Text-to-Speech~(TTS)~systems and propose the U-DiT architecture, exploring the potential of vision transformer architecture as the core component of the diffusion models in a TTS system. The modular design of the U-DiT architecture, inherited from the best parts of U-Net and ViT, allows for great scalability and versatility across different data scales.
The proposed U-DiT TTS system is a mel spectrogram-based acoustic model and utilizes a pretrained HiFi-GAN as the vocoder. The objective~(\ie Frechet distance)~ and MOS results show that our DiT-TTS system achieves state-of-art performance on the single speaker dataset LJSpeech.
Our demos are publicly available at: https://eihw.github.io/u-dit-tts/
\end{abstract}

\section{Introduction}
Text-to-speech (TTS) or speech synthesis systems are computer-based systems designed to convert text into spoken language \cite{shen2018natural}, which have been widely used in human–computer interaction~(HCI), most recently, for example, to power AI voice assistants. In recent years, dramatic breakthroughs in deep learning technology have resulted in the synthesis of more natural-sounding synthetic speech  \cite{shen2018natural,ren2020fastspeech,kim2020glow, popov2021grad, Triantafyllopoulos22-ASS, wang2023neural}. 

Modern Deep Neural Network~(DNN)-based TTS systems typically consist of two main components: an acoustic model and a vocoder \cite{shen2018natural, Triantafyllopoulos22-ASS} -- thus incorporating the two initial steps of a standard pipeline into one single model. 
The acoustic model converts the input text into time-frequency domain acoustic features, then, the vocoder synthesizes the raw waveforms conditioned on these acoustic features.
Since the introduction of Tacotron 2 \cite{shen2018natural}, the mel spectrogram has been the dominant acoustic feature in most modern TTS systems \cite{kim2020glow, ren2020fastspeech, popov2021grad}. 

Instead of directly generating mel spectrograms frame by frame in a sequence-to-sequence~(seq2seq)~architecture \cite{yasuda2021end}, generative models aim to learn the latent distribution of speech signals and sample from it to synthesize new ones. Variational Autoencoders~(VAEs) \cite{yasuda2021end}, Generative Adversarial Networks~(GANs) \cite{binkowski2019high}, and Flow-based generative models \cite{kim2020glow, valle2021flowtron} have produced more natural, coherent and faster speech synthesis utilizing their ability to capture the underlying relation between input text and speech.

Recently, Score-based Generative Models~(SGMs)~ \cite{song2019generative, song2020score, ho2020denoising, rombach2022high}, also known as Diffusion Probabilistic Models~(DPMs)~have achieved state-of-the-art performance in many research fields, such as image generation \cite{ho2020denoising,rombach2022high}, 3D shape creation \cite{zhou20213d}, natural speech synthesis \cite{popov2021grad,chen2022resgrad}, and music fragment generation \cite{mittal2021symbolic}.
SGMs are based on the simple but effective idea that a complex data distribution can be gradually turned into a simple one~(usually a normal distribution $N\sim(\sigma, I)$)~by iteratively adding scheduled noise. 
Meanwhile, a neural network can be trained to invert this procedure by following the trajectories of the reverse time forward process \cite{sohl2015deep}. 
In these SGMs, a modified U-Net architecture \cite{ho2020denoising} has proven to be well-suited as this diffusion model. Its ability to capture both local and global features with additional spatial self-attention blocks has lead to its widespread adoption in research fields. As a further improvement of the U-Net architecture, Diffusion Visual Transformer~(DiT) \cite{peebles2022dit} introduced the Visual Transformer~(ViT)~as the backbone of the diffusion model and achieved state-of-the-art image generation performance on the class-conditional ImageNet benchmarks.

 In this work, we propose a U-DiT architecture, thus exploring the possibility of ViT transforms as the core component in the backbone of a neutral diffusion-based TTS system. By combining the properties of the U-Net and ViT, both our objective results and MOS results show that U-DiT TTS generates higher-quality speech and more natural prosody compared to the state-of-the-art diffusion-based TTS system. 

\section{Score-based Generative Models}
\label{sec:sgms}
Score-based generative models~(SGMs) is a unified framework proposed by Song \et \cite{song2020score} which adopts a stochastic differential equation~(SDE) formulation \cite{kloeden1992stochastic}. They are a class of generative diffusion models that have the ability to learn the gradients of log probability density functions~($log PDF$ or score functions)~on a large number of noise-perturbed data distributions, and then generate samples with an iterative procedure called Langevin dynamics \cite{gparisi1981lngevin}, which is a class of Markov Chain Monte Carlo~(MCMC)~algorithms that generate samples from a probability distribution.

\subsection{SGMs with stochastic differential equations}
Instead of a finite number of noise distributions, the SGMs perturb the training data with a continuum of distributions that evolve over time according to a diffusion process, and the forward process $\{x_t\}_t$ can be described by an It\^{o} SDE \cite{steele2001stochastic}:
\begin{equation}
\label{equ:sde}
dx_t = f(x_t, t)dt + g(t)dw,
\end{equation}
where $f$ is $drift$ coefficient of $x(t)$, $g$ is the $diffusion$ coefficient of $x(t)$, and $w$ is a standard Wiener process~(or Brownian motion). Please note that the $t$ herein refers to the process step, and it has no relation to the time axis of the audio or its transformed version.

According to Anderson \cite{anderson1982reverse}, every SDE in the form of \autoref{equ:sde} should have a corresponding reverse SDE, which is also a diffusion process:
\begin{equation}
\label{equ:rvs_sde}
dx_t= \left[f(x_t, t) - g(t)^2\nabla_{x_t}log p_t(x_t)\right]dt + g(t)d\Bar{w}, 
\end{equation}

where $dt$ is an infinitesimal negative timestep and $\bar{w}$ is a standard Wiener process when $t$ reverses from $T$ to $0$. $\nabla_{x_t}log p_t(x_t)$ is the score function of a prior distribution $p(x)$ which can be approximated by a learned time-dependent score-based model $s_{\theta}$, such that $s_{\theta}(x_t, t) \approx \nabla_{x_t}log p_t(x_t)$.
\subsection{Diffusion Process in TTS systems}
\label{sec:sgm_fwd}

\textbf{Forward Process: }As mentioned above, a forward SDE diffusion process should be defined first to corrupt the input data to the prior distribution. In this work, we follow the forward diffusion process from the Grad-TTS \cite{popov2021grad}, in which the forward SDE is defined as:
\begin{equation}
\label{equ:fwd}
    dX_t = \frac{1}{2}\Lambda^{-1}(\mu - X_t)\beta_t dt + \sqrt{\beta_t}dW_t, \quad t\in[0, T], 
\end{equation}
where $\beta_t$ is the non-negative-valued noise schedule function, $\mu$ is vector, and $\Lambda$ is a diagonal matrix with positive elements.

Let $I$ be the identity matrix and the prior distribution can be defined as:

\begin{equation}
\label{equ:kernel}
p_{0t}\{X_t|X_0, \mu\} = \mathcal{N}(X_t; \rho(X_0, \Lambda, \mu, t), \delta(\Lambda, t)^2I)
\end{equation}

It is an Ornstein-Uhlenbeck SDE \cite{maller2009ornstein} and its mean $\rho(*)$ and variance $\delta(*)^2$ can be derived in a close form, according to the {\O}ksendal \cite{oksendal2003stochastic}, as:
\begin{equation}
\label{equ:fwd_mu}
\begin{aligned}
    \rho(X_0, \Lambda, \mu, t) = (I &- e^{-\frac{1}{2}\Lambda^{-1}\int^{t}_{0}\beta_sds})\mu \\&+ e^{-\frac{1}{2}\Lambda^{-1}\int^{t}_{0}\beta_sds}X_0
\end{aligned}
\end{equation}

\begin{equation}
\label{equ:fwd_delta}
\delta(\Lambda, t)I = \Lambda\left(I-e^{\Lambda^{-1}\int^{t}_{0}\beta_sds}\right)
\end{equation}

\textbf{Reverse Process:} From \autoref{equ:rvs_sde}, we can get the reverse SDE for the reverse diffusion process. However, to simplify and speed up the reverse procedure, Song \cite{song2020score} provides a probability flow ordinary differential equation~(ODE), which leads to the following reverse equation:
\begin{equation}
\label{equ:rev_sde}
dX_t=\left(\frac{1}{2}\Lambda^{-1}(\mu- X_t) - \nabla logp_t(X_t)\right)\beta_tdt
\end{equation}

\begin{figure}[t]
    \vspace{0.3cm}
    \hspace{-0.36cm}
    \includegraphics[width=0.51\textwidth]{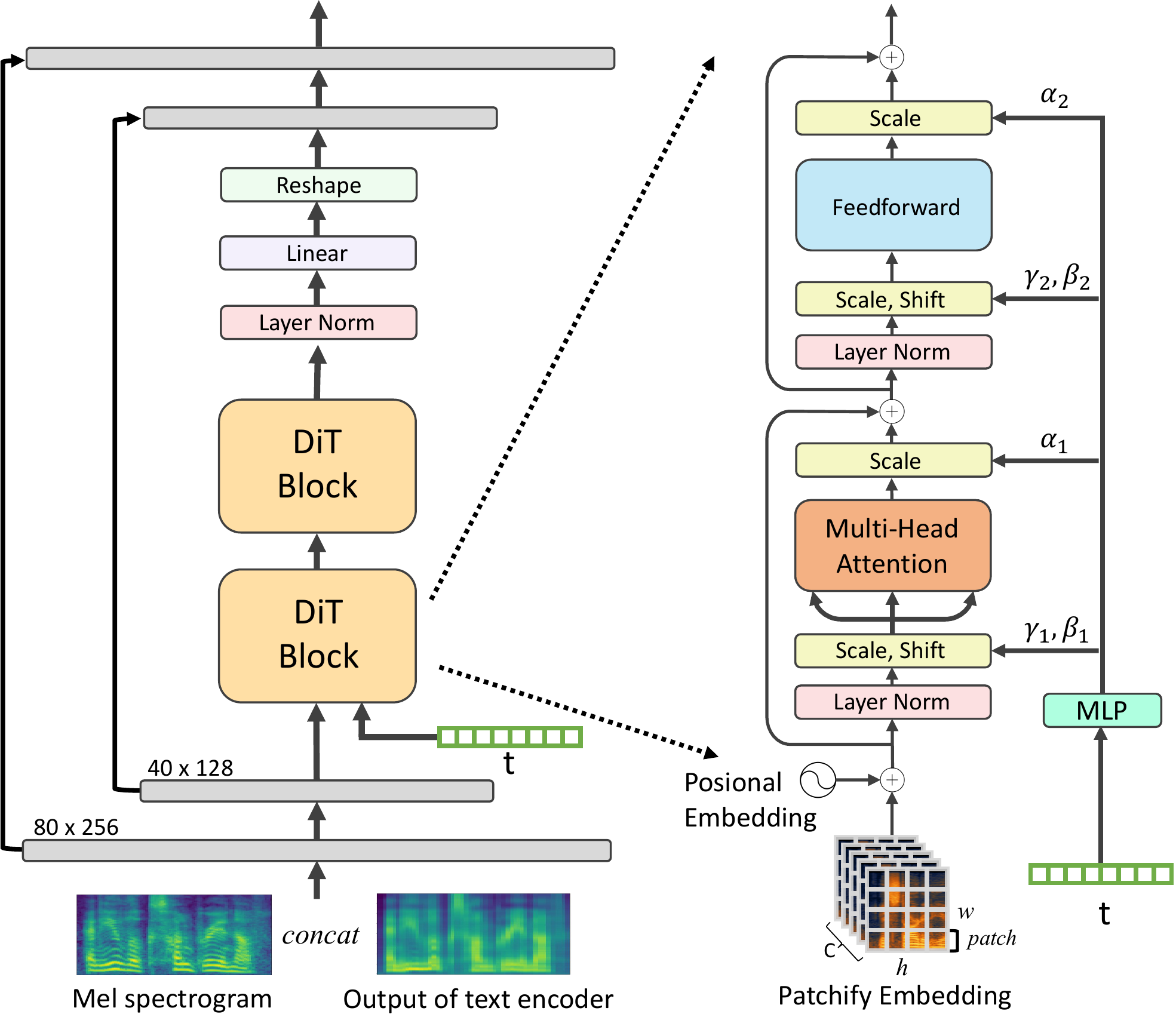}
    \caption{Overview of the U-DiT architecture. The downsampling part will transform the input into the latent space and the process step $t$, which is randomly sampled between $[0, 1]$ and encoded by the sinusoidal position embedding, will be applied as extra conditional information of the DiT blocks. At last, the upsampling part will convert the embeddings to the original size}
    \label{fig:U-DiT}
\end{figure}

\section{System Overview}
In this work, we propose the U-DiT architecture, which contains several diffusion visual transformer blocks, one of the ViT variants designed for diffusion process. This new architecture is designed to effectively capture both high-level semantic features and fine-grained details of mel spectrograms, leading to improved performance on speech synthesis tasks.

\subsection{DiT blocks}
The \autoref{fig:U-DiT}~(right)~demonstrates the detailed design of the proposed DiT block. The DiT block has retained the best practice of the ViT block design, while some minor but important modifications have been made to the standard ViT block.

\textbf{Patchify:} DiT will convert the input into a sequence of tokens like ViT. Then,  the standard frequency-based positional embedding will be applied to all tokens. We set the patch size to $[20, 64]$ in our model.

\textbf{Layer Normalization:} Normalizing a set of data points transforms them to a similar scale, which can improve generalization. In particular, adaptive layer normalization \cite{per2018film}~(adaLN)~enables smoother gradients and better generalization accuracy by normalizing the distributions of intermediate layers. It has been widely utilized by GAN architectures \cite{brock2018large} and in the standard diffusion U-Net backbone \cite{dhariwal2021diffusion}. Additionally, the diffusion U-Net backbone also employs a zero-initialization strategy for its final convolutional layers before residual connections, which has also been found to accelerate large-scale training on different tasks. The combination of the two techniques, adaLN-Zero, has proven to be a powerful design for further improving DiT performance and generalization on image synthesis \cite{peebles2022dit}. 

\textbf{Embedding MLP:} Since the diffusion process is very strongly time- and label-dependent~(if a label exists), a 4-layer multilayer perceptron~(MLP)~has been applied to regress the shift parameter $\gamma$ and $\beta$ from the sum of the time embedding vector $t_e$ and label embedding vector $l_e$. Furthermore, the dimension-wise scaling parameters $\alpha$, which are applied prior to any residual connections within the DiT block, are also regressed by the same MLP block.

\subsection{Latent space for U-}
In our experiments, we initially apply the DiT architecture as the backbone of the diffusion models in the high-resolution pixel space. However, this lead to a generated speech plagued by noise and lacking coherence. To address this issue, we first utilized the downsampling and upsampling components of the U-Net architecture from the DDPM \cite{ho2020denoising} to transform the input spectrograms into a latent space representation, which then served as the input of the DiT blocks. 
As shown in \autoref{fig:U-DiT}~(left), each layer in the downsampling part consists of multiple residual blocks with group normalization \cite{wu2018group} to extract the low-level features from the input, which are then processed by a following self-attention layer. 
Subsequently, the latent space features are partitioned into small patches and the spatial information of these patches is encoded by the sinusoidal position embedding. Finally, the upsampling part, which has a symmetrical structure to the downsampling part, restores the latent space features to their original size.

\subsection{Model Architechure}
We build our TTS system based on the structure from the Glow-TTS \cite{kim2020glow} and Grad-TTS \cite{popov2021grad}. Therefore, the U-DiT TTS system comprises three interrelated components: a text encoder, a duration predictor, and a decoder. 

\textbf{Text Encoder:}
The text encoder is composed of a pre-net, 6 transformer blocks with multi-head self-attention, and a final linear projection layer. The pre-net consists of 3 convolutional layers followed by a fully connected layer. This architecture is designed to effectively capture the meaningful features of the input text that can be used by the rest of the TTS system. 

\textbf{Duration predictor:} 
In this work, we follow the Glow-TTS \cite{kim2020glow} to apply an iterative approach called Monotonic Alignment Search~(MAS)~function to find an optimal monotonic surjective alignment between text and the corresponding speech. We also apply the duration predictor network from Fastspeech 2 \cite{ren2020fastspeech}, which is responsible for predicting the duration of each element of the input text. 

\textbf{Decoder:} As shown in \autoref{fig:U-DiT}, we employ the 2-layer U-Net downsampling and upsampling components in the decoder. We also test with different numbers of DiT blocks, including 2, 4, and 8. We observed that larger model sizes resulted in worse performance, potentially due to the limited amount of available data. Therefore, we settle on using 2 DiT blocks in our system. For the vocoder, we employ the HiFi-GAN \cite{kong2020hifi} to generate the final speech from the reconstructed mel spectrogram.

\section{Training Objectives}
Training objectives are crucial to guide the model's training towards the desired outcome for a given input. According to our model architecture, there are 3 different loss functions to jointly optimise our TTS system performance.

\textbf{Encoder loss:} in this study, we presume that the noise is generated from a Gaussian distribution with a diagonal covariance, which has proven to be effective in several TTS systems \cite{kim2020glow,popov2021grad, kim2022guided} by simplifying several mathematical derivations. Therefore,  we can set the text encoder output as an independent normal distribution $\boldsymbol{\Tilde{\mu}}_{A_{(i)}} \sim \mathcal{N}(\boldsymbol{\Tilde{\mu}}_{A_{(i)}}, I)$, which leads to a negative log-likelihood encoder loss:
\begin{equation}
\begin{aligned}
\label{equ:loss_enc}
    \mathcal{L}_{enc} = &-\sum^{N}_{i}log\varphi(y_i;\boldsymbol{\Tilde{\mu}}_{A_{(i)}}, I),
\end{aligned}
\end{equation}

where $\boldsymbol{\Tilde{\mu}}$ is the text encoder output, and $\varphi(*, \mu, I)$ is the $PDF$ of normal distribution $\mathcal{N}(\mu, I)$. To further improve effectiveness, we set the $\mathcal{L}_{enc}$ in the logarithmic domain, which makes the $\mathcal{L}_{enc}$ be the mean square error~(MSE)~loss.

\textbf{Duration loss:} we also train the Duration Predictor loss \cite{kim2020glow} $DP$ with the MSE loss in the logarithmic domain:
\begin{equation}
\label{equ:loss_dp}
    \begin{aligned}
        &d_i = \,log\sum^{N_{freq}}_{j=1} 1_{A^{*}(j)=i}, \qquad i=1,..,N_{freq} \\
              & \mathcal{L}_{DP}=MSE(DP(sg[\boldsymbol{\Tilde{\mu}}], d),
    \end{aligned}
\end{equation}
where $sg[*]$ is the stop gradient operator used to prevent $\mathcal{L}_{DP}$ affecting the text encoder parameters. With the stop gradient operation, it seems that the text encoder with duration predictor could be trained independently without a decoder. However, our results show that this only leads to an ineffective training process with a failure in alignment.

\textbf{Diffusion loss:} as mentioned in \autoref{sec:sgm_fwd}, our aim is to train a score model $s_\theta(x_t, t)$ to estimate gradients of log-density of the noisy data $X_t$. 
\begin{equation}
\label{equ:loss_diff_base}
    \mathcal{L}_t(X_0) = \mathbb{E}_{\epsilon_t}\left[\left\Vert s_\theta(X_t, t) + \lambda(\Lambda, t)^{-1}\epsilon_t\right\Vert_2^2\right],
\end{equation}
where $X_0$ is the target spectrogram sampled from the training data at time $t \in [0, T]$, which is sampled from a uniform distribution $\xi_t \sim \mathcal{N}(0, I)$ and according to \autoref{equ:kernel}, the noisy data $X_t$ can be sampled from:
\begin{equation}
\label{equ:loss_diff_xt}
    X_t = \rho(X_0, I, \mu, t) + \sqrt{\lambda_t}\xi_t.
\end{equation}
In \autoref{equ:loss_enc}, we already set the $\Lambda = I$, leading to a simplified noise distribution. The covariance matrix of the noise distribution is an identity matrix $I$ multiplied by a scalar:

\begin{equation}
\label{equ:loss_diff_cov}
    \lambda_t = 1 - e^{\int^{t}_{0}\beta_sds}.
\end{equation}

Finally, we have our diffusion loss function as follows:
\begin{equation}
\label{equ:loss_diff}
    \mathcal{L}_{diff} = \mathbb{E}_{X_0, t}\left[\lambda_t \mathbb{E}_{\xi_t}\left[\left\Vert s_{\theta}(X_t, \mu, t)+ \frac{\xi_t}{\sqrt{\lambda_t}} \right\Vert^2_2\right]\right].
\end{equation}

\section{Experiments}
\subsection{Dataset}
In this work, we used the standard LJSpeech dataset \cite{ljspeech17} with the official training and test set. The dataset consists of 13,100 audio files with a total duration of approximately 24 hours and is accompanied by transcriptions of the audio in the form of plain text files. The audio is encoded in the WAV format at a sampling rate of 22.05 kHz. 

\subsection{Experiment settings}
For audio feature generation, we used 80-dimensional mel spectrograms with a 1024 window length and a 256 hop length. The input text was converted to phonemes through CMU pronouncing dictionary. The output of the text encoder was concatenated with the mel spectrograms. Then, we randomly took 256 frame lengths on the time axis of speech representation as the final input.

For the diffusion process, we used the same noise schedule as the Grad-TTS \cite{popov2021grad} $\beta_t = 0.05 + (20 - 0.05)t, \,t\in[0, 1]$. During the training phase, we used the Adam optimizer with a learning rate of $1e-4$ and the batch size is 32. We also applied gradient clipping to prevent exploding gradients. All of our models were trained and tested in a Python 3.9.12 and PyTorch 1.12.1 environment using an Nvidia RTX 3080 GPU.

The objective evaluation metrics we apply in this work are Frechet Distance~(FD), Log-spectral Distance~(LSD), and Kullback–Leibler Divergence~(KLD). The FD \cite{frechet1906quelques} measures the similarity between generated speech samples and the target speech samples. 
The log-spectral distance (LSD) \cite{gray1976distance}, also referred to as log-spectral distortion, is a distance measure between two spectra.
The KLD \cite{kullback1951information} measures dissimilarity between probability distributions.  All the objective metrics are from AudioLDM\footnote[1]{https://github.com/haoheliu/audioldm\_eval} \cite{liu2023audioldm} and all the speech samples are resampled to 16KHz as the evaluation system requested.

 To further evaluate the performance of the proposed TTS system, Mean Opinion Score~(MOS)~tests were conducted. In the MOS study, $14$ samples were selected from each of the $4$ systems listed in \autoref{tab:metric} and were annotated by $16$ individuals. Specifically, for each source text, the subjects were asked to rate the overall quality (\eg naturalness, intelligibility) of the $5$ corresponding audios shuffled in a five-point Likert score ($5$: Excellent, $4$: Good, $3$: Fair, $2$: Poor, $1$: Bad). The subset of the samples used in the MOS study is available on the demo page.

\subsection{Inference}

During inference, we segmented the input text to address the input size restrictions of the transformer architecture. We adjusted the text phonemes segmentation length between 22 and 25 since we find the output length of the text encoder was slightly less than 256 when the total phonemes count is in this range. Meanwhile, instead of sampling $x_T$ from  the $N(\mu, I)$, we applied a hyperparameter $\tau$ and started to sample from the distribution $N(\mu, I/\tau)$. In our following experiments, we opted for the $reverse\_step = 80$ and temperature $\tau=1.5$.
\subsection{Results}

We compared the performance of our models on the objective evaluation with 1) Grad-TTS \cite{popov2021grad} with the official implementation as our baseline system, 2) GT\textit{mel}~(mel + HiFi-GAN), where we first converted the ground truth speech into mel spectrogram and then used HiFi-GAN to convert it back. 

\begin{table}[ht]
\centering
\caption{Performance comparison on LJSpeech test set}
    \begin{tabular}{l c c c |c}
        \toprule
        Metric   & FD$\downarrow$ & LSD$\downarrow$ & KLD$\downarrow$ & MOS\\ \hline
        GT      &  -        & -      & -      & 4.10    \\ 
        GT\textit{mel}      &  0.6985        & 1.8202      & 0.0109  &     3.90    \\\hline
        Grad-TTS \cite{popov2021grad}      & 3.0046         & 2.1765       & 0.0647 &     3.62     \\ 
 
        U-DiT TTS     & \textbf{0.8960}        & \textbf{2.1745}   & \textbf{0.0316}    &     \textbf{3.91} \\
        \bottomrule
    \end{tabular}
\label{tab:metric}
\end{table}

\autoref{tab:metric} shows that the proposed U-DiT TTS systems outperform the baseline Grad-TTS system. The FD metric, which indicates speech quality, is lower than Grad-TTS by a large margin while close to the GT\textit{mel}, which is also the target in the training phase. KLD and LSD also show that our systems have a better similarity to the ground truth on mel spectrogram. All the results on the objective evaluation indicate that the U-DiT is better capable of capturing the spatial relationships between different parts of the input data when compared to the U-Net. The MOS scores also demonstrate the U-DiT TTS outperforms the Grad-TTS by 0.29, which means that it can synthesize more natural speech. Furthermore, the MOS score of U-DiT TTS indicates a competitive synthesis quality compared to the GT\textit{mel} but is still marginal to the GT audio. We observe that the pre-trained HiFi-GAN model is unable to fully restore the original audio and U-DiT TTS produces the mel spectrogram with less background noise and system noise. In the subjective MOS study, participants found the synthesized audio to be slightly clearer and easier to understand, resulting in a higher MOS score. 

\subsection{Ablation study}
To find the optimal settings for the best synthesis quality, we performed a parameter ablation on the reverse step and temperature in the reverse process. We used FD and KLD as the primary evaluation metrics in the ablation study since they share the same embeddings extracted by CNN14 from PANNs \cite{kong2020panns}.

\begin{figure}[htbp]
    \hspace{-0.16cm}
    \includegraphics[width=0.48\textwidth]{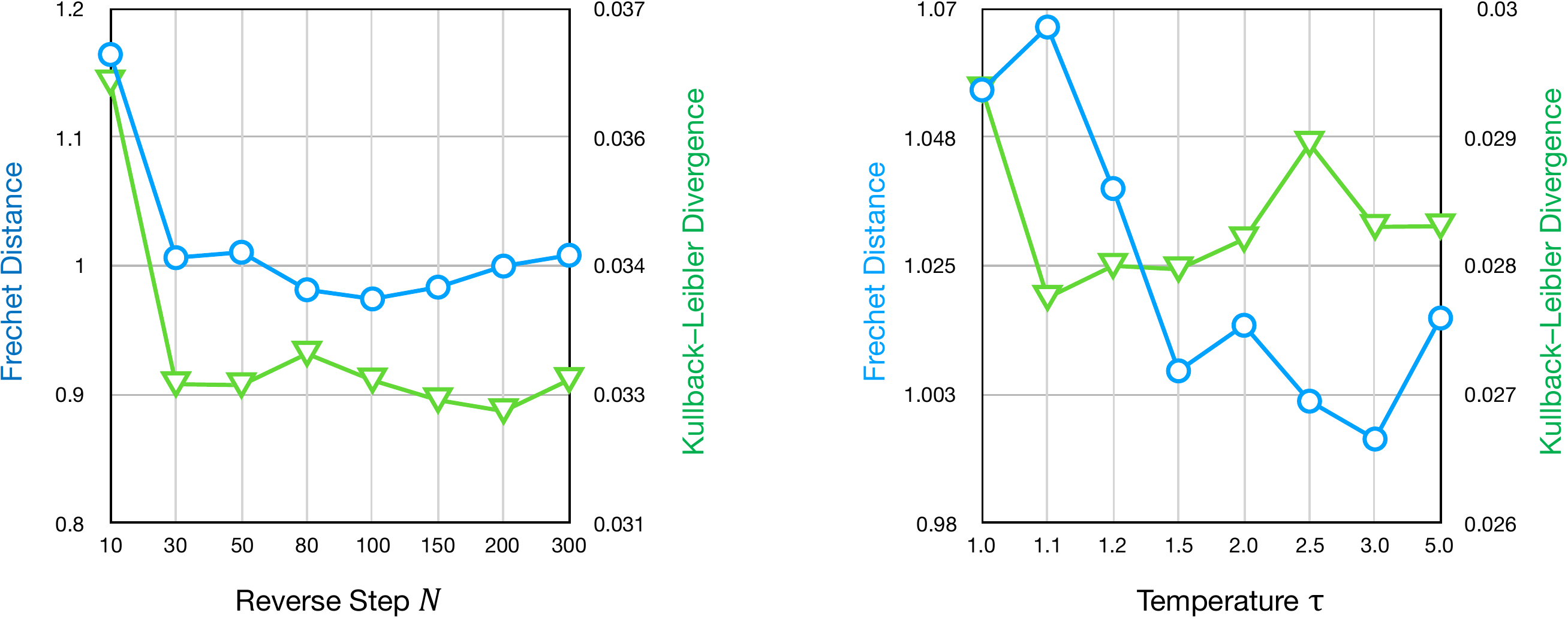}
    \caption{Synthesis Performance comparison on different parameters. Synthesis quality in FD and LSD metric as a function of the reverse steps and temperature $\tau$ in the reverse SDE}
    \vspace{-0.5cm}
    \label{fig:aba}
\end{figure}

The reverse steps of the SGMs can have a great impact on the final generation quality \cite{ho2020denoising}. In \autoref{fig:aba} left, we show the FD and KLD metrics as a function of the reverse steps. Our results show that a good tradeoff between high quality and efficiency be achieved with 80 to 150 steps. Our experiments also show that a reverse step of 30 achieves acceptable speech quality while maintaining a shorter synthesis time.

While the larger step size $1/N$ would negatively impact the quality of the output mel spectrograms, adjusting the temperature $\tau$ can help mitigate this effect and improve the generation quality without affecting the inference time. With the reverse step fixed to 80, we explore the effect of temperature $\tau$ on our system. According to \autoref{fig:aba} right, our experiments demonstrate that the optimal range for $tau$ is between 1.1 and 2. Within this range, we observed a consistent improvement in the quality of the generated speech, as indicated by the continuous decrease in FD and a marginal growth on KLD. 

\section{Conclusion}
In this work, we proposed U-DiT architecture exploring the possibility of ViT transforms as the core component in the backbone of the neutral diffusion-based TTS system. The results show that the U-DiT TTS system can effectively generate higher-quality natural speech compared to the state-of-the-art diffusion-based TTS system. However, there are still some limitations, such as the fixed input size, and the high requirement for the quality of training data. In future work, we plan to address the limitations and further optimize the system to achieve better performance.

Overall, the U-DiT architecture provides a promising solution for the TTS task, with the potential to significantly advance the state of the art in this field.

\section{Acknowledgement}
This work was funded by the China Scholarship Council~(CSC), Grant \#\,202006290013, and by the DFG (German Research Foundation),  Reinhart Koselleck-Project AUDI0NOMOUS (Grant No.\ 442218748).
\bibliographystyle{IEEEtran}
\bibliography{sample}

\end{document}